\documentclass[12pt]{article}
\usepackage{amsmath}
\input{epsf}
\def\ltap{\raisebox{-.4ex}{\rlap{$\,\sim\,$}} \raisebox{.4ex}{$\,<\,$}}
\def\gtap{\raisebox{-.4ex}{\rlap{$\,\sim\,$}} \raisebox{.4ex}{$\,>\,$}}

\newcommand\as{\alpha_{\mathrm{S}}}
\newcommand\f[2]{\frac{#1}{#2}}

\def\dO{{\cal D}_{0}}
\def\dl{{\cal D}_{1}}
\def\dll{{\cal D}_{2}}
\def\dlll{{\cal D}_{3}}

\def\beq{\begin{equation}}
\def\eeq{\end{equation}}
\def\beeq{\begin{eqnarray}}
\def\eeeq{\end{eqnarray}}

\def\to{\rightarrow}
\def\nn{\nonumber}

\def\ms{${\overline {\rm MS}}$}

\def\sqr{\sqrt{1 - \pitcut^2 }}
\def\pitcut{\pi_T}

\begin{document}
\begin{titlepage}
\renewcommand{\thefootnote}{\fnsymbol{footnote}}
\begin{flushright}
    CERN--TH/2001-222   \\ hep-ph/0111164
     \end{flushright}
\par \vspace{10mm}

\begin{center}
{\Large \bf
Direct Higgs production and jet veto \\
at the Tevatron and the LHC in 
NNLO QCD\footnote{This work was supported in part 
by the EU Fourth Framework Programme ``Training and Mobility of Researchers'', 
Network ``Quantum Chromodynamics and the Deep Structure of
Elementary Particles'', contract FMRX--CT98--0194 (DG 12 -- MIHT).}
\\[1.ex]
}

\end{center}
\par \vspace{2mm}
\begin{center}
{\bf Stefano Catani${}^{(a)}$~\footnote{On leave of absence 
from INFN, Sezione di Firenze, Florence, Italy.}, Daniel de Florian${}^{(b)}$~\footnote{Partially supported by CONICET and Fundaci\'on Antorchas} }
\hskip .2cm
and
\hskip .2cm
{\bf Massimiliano Grazzini${}^{(c,d)}$}\\

\vspace{5mm}

${}^{(a)}$Theory Division, CERN, CH-1211 Geneva 23, Switzerland \\

${}^{(b)}$ Departamento de F\'\i sica, FCEYN, Universidad de Buenos Aires,
(1428) Pabell\'on 1 Ciudad Universitaria, Capital Federal, Argentina  
 \\

${}^{(c)}$ Dipartimento di Fisica, Universit\`a di Firenze, I-50019 Sesto Fiorentino, Florence, Italy\\

${}^{(d)}$INFN, Sezione di Firenze, I-50019 Sesto Fiorentino, Florence, Italy

\vspace{5mm}

\end{center}

\par \vspace{2mm}
\begin{center} {\large \bf Abstract} \end{center}
\begin{quote}
\pretolerance 10000

We consider 
Higgs boson production through gluon--gluon
fusion in hadron collisions, when a veto is applied on
the transverse momenta of the accompanying hard jets.
We compute the QCD radiative corrections to this process at NLO and NNLO.
The NLO calculation is complete. The NNLO calculation uses the
recently evaluated NNLO soft and virtual QCD contributions to the inclusive
cross section.
We find that the jet veto reduces the impact
of the NLO and NNLO contributions, the reduction being more
sizeable
at the LHC than at the Tevatron.

\end{quote}

\vspace*{\fill}
\begin{flushleft}
     CERN--TH/2001-222 \\ November 2001 

\end{flushleft}
\end{titlepage}

\setcounter{footnote}{1}
\renewcommand{\thefootnote}{\fnsymbol{footnote}}


\section{Introduction}

\label{sec:intro}

The mechanism of electroweak symmetry breaking is one of the main issues 
in high-energy particle physics. 
In the Standard Model (SM) and in its supersymmetric extensions the mechanism is accomplished by elementary scalar doublets. 
The Higgs boson(s) \cite{Gunion:1989we}
are thus a fundamental ingredient of the theory,
and their search is the main goal of current and future colliders.

The direct search at LEP implies a lower limit of
$M_H > 114.1$~GeV (at $95\%$ CL) \cite{lep}
on the mass $M_H$ of the SM Higgs boson, and shows an excess of events, which may
indicate the production of a Higgs boson with mass near 115~GeV 
\cite{lepc,leppapers,lep}.
Global SM fits to electroweak precision measurements favour a light Higgs
($M_H \ltap 200$~GeV) \cite{ew}.

After the end of the LEP era, the search for the Higgs boson 
will be carried out at hadron colliders. Depending on the luminosity
delivered to the CDF and D0 detectors during the forthcoming Run II, the 
Tevatron experiments can yield evidence for a Higgs boson with 
$M_H \ltap 180$~GeV and may be able to discover (at the $5\sigma$ level)
a Higgs boson with $M_H \ltap 130$~GeV \cite{Carena:2000yx}.
At the LHC, the SM Higgs boson can be discovered over the full mass range up
to $M_H \sim 1$~TeV after a few years of running \cite{atlascms}.

The dominant mechanism for SM Higgs boson production at hadron colliders is
gluon--gluon fusion
through 
a heavy-quark (top-quark) loop \cite{Georgi:1978gs}.
This mechanism is often called {\em direct} 
Higgs production, to distinguish it from {\em associated} production of Higgs 
boson and vector bosons, heavy quarks, jets, and so forth.

At the LHC \cite{Spira:1998dg}, 
$gg$ fusion exceeds all the other production
channels by a factor decreasing from 8 to 5 when $M_H$ increases from 100 to
200~GeV. When $M_H$ approaches 1~TeV, $gg$ fusion still provides about 
$50\%$ of the total production cross section.
At low values of $M_H$ the dominant decay mode is $H\to b{\bar b}$, 
but it is swamped by the QCD background. This decay mode is thus exploited
\cite{atlascms}
in the low-mass range 100~GeV~$\ltap M_H \ltap$~120~GeV only in the case of
the associated production $Ht{\bar t}$. In the case of direct production,
the most important decay channels for Higgs searches \cite{atlascms}
are the rare decay 
$H\to \gamma\gamma$ in the low-mass range 100~GeV~$\ltap M_H \ltap$~140~GeV,
and the channel $H\to ZZ \to 4l$ in the mass range 
130~GeV~$\ltap M_H \ltap$~700~GeV. In the intermediate-mass region
150~GeV~$\ltap M_H \ltap$~190~GeV, direct Higgs production followed by the 
decay $H\to WW\to l^+l^-\nu {\bar \nu}$ is also relevant.
The strong angular correlations of the final-state leptons from $W$ decay
are an important ingredient for this discovery channel
\cite{angcor}.

At the Tevatron, $gg$ fusion
remains 
the main production channel
\cite{Carena:2000yx}, but it is experimentally less important than
at the LHC because
the decay rate $H\to \gamma\gamma$
is too low to be observed.
When $M_H\ltap 135$~GeV, the most promising discovery mechanism 
\cite{Carena:2000yx} is thus 
the associated production ($q {\bar q} \to V^* \to H V$)
of the Higgs boson with a vector boson $V$ ($V=W$ or $Z$), whose leptonic decay
provides the trigger for the signal from $H \to b{\bar b}$.
Nevertheless,
since the decays $H\to W^*W^*, Z^*Z^*$ have
increasing branching fractions for $M_H\gtap 130$ GeV, 
$gg\to H\to W^*W^*,Z^*Z^*$ are the natural channels
to consider when 140~GeV~$\ltap M_H \ltap$~180~GeV. In particular, 
the decay mode $W^*W^*\to l^+l^-\nu {\bar \nu}$ is quite important
\cite{Carena:2000yx},
since it is cleaner than $W^*W^*\to l\nu jj$, and the decay rate $H\to W^*W^*$
is higher than $H\to Z^*Z^*$ by about one order of magnitude.

An important background for the direct Higgs signal 
$H\to W^*W^*\to l^+l^-\nu {\bar \nu}$
is $t {\bar t}$ production ($tW$ production is also important at the LHC), where
$t \to l{\bar \nu} b$, thus leading to $b$ jets with high $p_T$ in the final 
state. If the $b$ quarks are not identified, 
a veto cut on the transverse momenta of the jets accompanying
the final-state leptons
can be applied to enhance the signal/background ratio.
Imposing a jet veto turns out to be essential, both at
the Tevatron \cite{Carena:2000yx,Han:1999ma} and at the LHC \cite{atlascms},
to cut the hard $b$ jets arising from this background process.

The next-to-leading order (NLO) QCD corrections to $gg$ fusion are large
\cite{Dawson:1991zj, Djouadi:1991tk, Spira:1995rr}. Approximate evaluations 
\cite{Kramer:1998iq} of higher-order terms suggest that their effect can still
be sizeable. The computation of the next-to-next-to-leading order (NNLO)
corrections is thus important to better estimate the cross section for direct
Higgs production.

Recently two 
groups~\cite{Catani:2001ic,Harlander:2001is} have performed a first step
in this direction, by evaluating the soft and 
virtual contributions to the NNLO partonic cross section 
${\hat \sigma}(gg\to H+X)$ in the large-$M_{\rm top}$ approximation.
Our calculation~\cite{Catani:2001ic} was performed by combining Harlander's
result~\cite{Harlander:2000mg} for the two-loop amplitude $gg\to H$
with the soft factorization formulae for tree-level~\cite{Campbell:1998hg,Catani:2000ss} 
($gg \to Hgg, Hq{\bar q}$)
and one-loop~\cite{Bern:1998sc,Catani:2000pi} ($gg \to Hg$) amplitudes,
and then using the technique of Ref.~\cite{Matsuura:1989sm}.
The independent calculation of Ref.~\cite{Harlander:2001is} used a different
method, and the analytical results fully agree.

In Ref.~\cite{Catani:2001ic}
we also studied the quantitative impact of the soft and virtual NNLO 
contributions on direct Higgs boson production at  the LHC.
This was consistently done by using the recent MRST2000 set of parton distribution functions 
\cite{mrst2000}, which includes (approximated \cite{vnvogt}) NNLO densities.
In this paper we first perform
an analogous study\footnote{Part of these results 
was anticipated in Ref.~\cite{moriond}.} at the Tevatron Run II, and we show
that the soft and virtual contributions are expected to give a
fairly good estimate of the complete NNLO result.

In the second part of the paper we study the effect of a jet veto
on direct Higgs production beyond the leading order (LO).   
We present new NLO and NNLO calculations
for the vetoed cross section
both at the Tevatron and at the LHC.
These calculations are performed by subtracting
the LO (NLO) cross
section for the production of Higgs plus jets
from the corresponding inclusive NLO (NNLO) cross section.
The LO and NLO subtracted cross section for the production of Higgs plus jets
is evaluated in the large-$M_{\rm top}$ limit, but without any soft and virtual
approximations. At LO we derive analytical expressions, while at the NLO 
we use the numerical program of Ref.~\cite{deFlorian:1999zd}.

The paper is organized as follows.
In Sect.~\ref{sec:theo} we recall the theoretical framework and the
analytical results for the soft and virtual
NNLO corrections. In Sect.~\ref{pheno} we present the results for inclusive Higgs
production at the Tevatron Run II. In Sect.~\ref{sec:veto} we present the results obtained
by applying a jet veto on the inclusive cross section,
both at the Tevatron Run II and at the LHC.
Our conclusions are summarized in Sect.~\ref{sec:conc}.

\section{Inclusive QCD cross section at NNLO}
\label{sec:theo}

We consider the collision of two hadrons $h_1$ and $h_2$ with 
centre-of-mass energy ${\sqrt s}$. The inclusive cross section for the 
production of the SM Higgs boson  can be written 
as
\begin{align}
\label{had}
\sigma(s,M_H^2) =& 
\sum_{a,b} \int_0^1 dx_1 \;dx_2 \; f_{a/h_1}(x_1,\mu_F^2) 
\;f_{b/h_2}(x_2,\mu_F^2) \int_0^1 dz \;\delta\!\left(z -
\frac{\tau_H}{x_1x_2}\right) \nn \\
& \cdot \sigma_0\,z\;G_{ab}(z;\as(\mu_R^2), M_H^2/\mu_R^2;M_H^2/\mu_F^2) \;,
\end{align}
where $\tau_H=M_H^2/s$, and $\mu_F$ and $\mu_R$ are factorization and 
renormalization scales, respectively. 
The parton densities of the colliding hadrons are denoted by 
$f_{a/h}(x,\mu_F^2)$ and the subscript $a$ labels the type
of massless partons ($a=g,q_f,{\bar q}_f$,
with $N_f$ different flavours of light quarks). 
We use parton densities as defined in the \ms\ factorization scheme.

From Eq.~(\ref{had}) the cross section ${\hat \sigma}_{ab}$ for the partonic 
subprocess $ab \to H + X$ at the centre-of-mass energy 
${\hat s}=x_1x_2s=M_H^2/z$ is
\begin{equation}
\label{sigmapart}
{\hat \sigma}_{ab}({\hat s},M_H^2) = \frac{1}{\hat s} 
\;\sigma_0 M_H^2 \;G_{ab}(z) = \sigma_0 \;z \;\;G_{ab}(z) \;,
\end{equation}
where the term $1/{\hat s}$ corresponds to the flux factor and leads to 
an overall $z$ factor. The  Born-level cross section $\sigma_0$ and 
the hard coefficient function $G_{ab}$ arise 
from the phase-space integral of the matrix elements squared. 

Equation~(\ref{had}) can also be recast in the form
\begin{equation}
\label{sighad}
\sigma(s,M_H^2)=\sigma_0\,\tau_H
\sum_{a,b}\int_{\tau_H}^1 \f{d\tau}{\tau}\, {\cal L}_{ab/h_1h_2}(\tau,\mu_F^2)\, 
G_{ab}(\tau_H/\tau) \;,
\end{equation}
where the function
\begin{equation}
\label{lumin}
{\cal L}_{ab/h_1h_2}(\tau,\mu_F^2)=\int_{\tau}^1 \f{dx}{x}
f_{a/h_1}(x,\mu_F^2) f_{b/h_2}(\tau/x,\mu_F^2)  
\end{equation}
is the parton luminosity that weights the event initiated by partons $a$ and
$b$.

The incoming partons $a,b$ couple to the Higgs boson
through heavy-quark loops and, therefore,  $\sigma_0$ and $G_{ab}$ also depend on the 
masses $M_Q$ of the heavy quarks. The Born-level contribution $\sigma_0$
is~\cite{Georgi:1978gs}
\begin{equation}
\label{borncs}
\sigma_0=\f{G_F}{288\pi\sqrt{2}} 
\;\left| \sum_Q A_Q\!\left(\frac{4M_Q^2}{M_H^2}\right) \right|^2 \;\;,
\end{equation}
where $G_F=1.16639 \times 10^{-5}$ GeV$^{-2}$ is the Fermi constant,
and the amplitude $A_Q$ is given by
\begin{align}
A_Q(x)  &= \f{3}{2} x \Big[ 1+(1-x) f(x) 
\Big] \;,\nonumber \\
f(x) & = \left\{ \begin{array}{ll}
\displaystyle \arcsin^2 \frac{1}{\sqrt{x}} \;, & x \ge 1 \\
\displaystyle - \frac{1}{4} \left[ \ln \frac{1+\sqrt{1-x}}
{1-\sqrt{1-x}} - i\pi \right]^2 \;, & x < 1
\end{array} \right. \;.
\end{align}
In the following we limit ourselves to considering
the case of a single heavy quark, the top quark, and $N_f=5$ light-quark
flavours. We always use $M_{\rm top}$ ($M_{\rm top}=176$~GeV) 
 to denote the on-shell pole mass of the top quark.

The coefficient function $G_{ab}$ in Eq.~(\ref{had}) is computable in QCD
perturbation theory according to the expansion
\begin{align}
\label{expansion}
&G_{ab}(z;\as(\mu_R^2), M_H^2/\mu_R^2;M_H^2/\mu_F^2) =
\as^2(\mu_R^2) \sum_{n=0}^{+\infty} \left(\f{\as(\mu_R^2)}{\pi}\right)^n
\;G_{ab}^{(n)}(z;M_H^2/\mu_R^2;M_H^2/\mu_F^2) \nn \\
&= \as^2(\mu_R^2)
G_{ab}^{(0)}(z) + \f{\as^3(\mu_R^2)}{\pi} \;
G_{ab}^{(1)}\left(z;\frac{M_H^2}{\mu_R^2};\frac{M_H^2}{\mu_F^2}\right) 
+ \f{\as^4(\mu_R^2)}{\pi^2} \;
G_{ab}^{(2)}\left(z;\frac{M_H^2}{\mu_R^2};\frac{M_H^2}{\mu_F^2}\right) 
+ {\cal O}(\as^5) \;,
\end{align}
where the (scale-independent) LO contribution is
\begin{equation}
G_{ab}^{(0)}(z) = \delta_{ag} \; \delta_{bg} \;\delta(1-z) \;.
\end{equation}

Throughout the paper we work 
in the framework of the large-$M_{\rm top}$ expansion,
where one can exploit the effective-lagrangian approach
\cite{efflag,Chetyrkin:1997iv,Kramer:1998iq}
to embody the heavy-quark loop in an effective vertex.
However,
unless otherwise stated, we include in $\sigma_0$ the full dependence on $M_{\rm top}$.
This approximation \cite{Spira:1995rr,Kramer:1998iq}
turns out to be very good when $M_H \leq 2M_{\rm top}$, and it is 
still accurate\footnote{The accuracy of this approximation
when $M_H \ltap 2M_{\rm top}$ may not be accidental. In fact, as discussed in
the following, the main part of the QCD corrections to direct Higgs production
is due to parton radiation at relatively low transverse momenta. Such radiation
is weakly sensitive to the mass of the heavy quark in the loop.}
to better than $10\%$ when $M_H \ltap 1$~TeV.

The NLO coefficients $G_{ab}^{(1)}(z)$ are known
\cite{Dawson:1991zj,Spira:1995rr}. Their explicit expressions (see e.g. 
Ref.~\cite{Catani:2001ic}) contain
three kinds of contributions:
\begin{itemize}
\item Virtual and soft contributions, 
which are respectively proportional to $\delta(1-z)$ and to the distributions
${\cal D}_i(z)$, 
where
\begin{equation}
{\cal D}_i\equiv \left[\f{\ln^i(1-z)}{1-z}\right]_+\, .
\end{equation}
These are the most singular terms when $z\to 1$.
\item Purely collinear logarithmic contributions of the type $\ln^i(1-z)$.
These contributions give the next-to-dominant singular terms
when $z\to 1$.
\item Hard contributions, which are finite in the limit $z\to 1$.
\end{itemize}

The soft-virtual (SV) approximation is defined \cite{Catani:2001ic} by keeping 
only the terms proportional to $\delta(1-z)$ and ${\cal D}_i(z)$
in the coefficient $G_{ab}$.
In this approximation only the 
$gg$ channel contributes and
we have
\begin{align}
\label{gg1SV}
G_{ab}^{(1) {\rm SV}}(z;M_H^2/\mu_R^2;M_H^2/\mu_F^2)&= \delta_{ag} \delta_{bg}
\left[ \delta(1-z) \left( \f{11}{2} + 6 \zeta(2) + 
\f{33-2N_f}{6}\ln\f{\mu_R^2}{\mu_F^2}\right) \right.\nn\\
&\left. +6 \dO\, \ln\f{M_H^2}{\mu_F^2} +12\,\dl \right] \;.
\end{align}

The NNLO coefficients $G_{ab}^{(2)}$ are not yet completely known,
but their SV approximation has been computed
in Refs.~\cite{Catani:2001ic,Harlander:2001is}. It reads:
\begin{align}
\label{gg2SV}
G_{gg}^{(2){\rm SV}}(z;&M_H^2/\mu_R^2,M_H^2/\mu_F^2)= \delta(1-z)
\Bigg[ \f{11399}{144} +\f{133}{2}\zeta(2) -\f{9}{20}\zeta(2)^2
  -\f{165}{4}\zeta(3)\nn\\ 
&\hspace{3.1cm}
+  \left(\f{19}{8}+\f{2}{3} N_f\right)
 \ln\f{M_H^2}{M_{\rm top}^2}
  +N_f \left( -\f{1189}{144} -\f{5}{3} \zeta(2) +\f{5}{6}\zeta(3)\right)
 \nn \\
&\hspace{3.1cm}
+\f{\left(33-2N_f\right)^2}{48}\ln^2\f{\mu_F^2}{\mu_R^2}- 18\,\zeta(2)
\ln^2\f{M_H^2}{\mu_F^2}\nn \\
&\hspace{3.1cm}
+ \left( \f{169}{4}+\f{171}{2}\zeta(3)- \f{19}{6}N_f
 +\left(33-2N_f\right)\zeta(2)\right)
\ln\f{M_H^2}{\mu_F^2}\nn \\
&\hspace{3.1cm}
+\left(-\f{465}{8}+\f{13}{3}N_f-\f{3}{2}\left(33-2N_f\right)\zeta(2)\right)
\ln\f{M_H^2}{\mu_R^2}
\Bigg] \nn \\
&
+\dO \Bigg[ -\f{101}{3} + 33 \zeta(2) + \f{351}{2} \zeta(3)+N_f\left(\f{14}{9}-2\zeta(2)\right)+
\left( \f{165}{4}-\f{5}{2}N_f \right)\ln^2\f{M_H^2}{\mu_F^2}\nn\\
&\hspace{1.2cm}
-\f{3}{2}\left(33-2N_f\right)\ln\f{M_H^2}{\mu_F^2}\,\ln\f{M_H^2}{\mu_R^2}  
+ \left( \f{133}{2} - 45 \zeta(2) - \f{5}{3} N_f \right)\ln\f{M_H^2}{\mu_F^2}
\Bigg] \nn \\
&
+\dl \Bigg[ 133 -90 \zeta(2) - \f{10}{3} N_f 
+ 36 \ln^2\f{M_H^2}{\mu_F^2}
+\left(33 -2 N_f \right)\left(2\ln\f{M_H^2}{\mu_F^2}-3\ln\f{M_H^2}{\mu_R^2}\right)
\Bigg]\nn \\
&+\dll \left[ -33+2 N_f +108 \ln\f{M_H^2}{\mu_F^2}\right] \nn \\
&+ 72\, \dlll \;\;.
\end{align}

To give a rough idea of the numerical hierarchy of the LO, NLO and NNLO 
contributions, we can set $\mu_F=\mu_R=M_H=115$~GeV and $\as(M_H)=0.112$.
The first three terms in the expansion (\ref{expansion}) thus give
\begin{align}
G_{gg}^{\rm SV}(z)=\as^2\Big\{
\delta(1-z)&\left(\;1\; + \; 0.548 \; + \; 0.105\;\right)\nn\\
    +\dO   &\left(\;0\; + \; ~~~~~0\; + \; 0.283\;\right)\nn\\
    +\dl   &\left(\;0\; + \; 0.428 \; - \; 0.040\;\right)\nn\\
    +\dll  &\left(\;0\; + \; ~~~~~0\; - \; 0.029\;\right)\nn\\
    +\dlll &\left(\;0\; + \; ~~~~~0\; + \; 0.092\;\right)\;\Big\}\nn \;\;.
\end{align}

Since $1 \geq z = \tau_H/\tau \geq \tau_H$,
the SV terms are certainly the dominant contributions to the cross section
in the kinematic region near the hadronic threshold $(\tau_H = M_H^2/s \sim 1)$. At fixed
$s$, this means that these terms certainly dominate in the case of heavy Higgs
bosons. However, the SV terms  
can give the dominant effect even long before the
threshold region in the hadronic cross section is actually approached
\cite{Contogouris:1990zq}. 
This is because, in the evaluation of the hadronic cross section
in Eq.~(\ref{sighad}), the partonic cross section $\hat \sigma_{ab}({\hat
s},M_H^2)$ has to be weighted with the parton luminosities,
which are strongly suppressed at large $\tau={\hat s}/s=x_1 x_2$. In other
words, owing to the strong suppression of the parton densities $f_a(x,\mu_F^2)$
at large $x$, the partonic centre-of-mass energy $\sqrt {\hat s}$ is typically
substantially smaller than $\sqrt s$ ($\langle {\hat s} \rangle = 
\langle x_1 x_2 s \rangle = \langle \tau \rangle s$), and the variable 
$z=M_H^2/\hat s$ in
$G_{ab}(z)$ can be close to unity also when $\sqrt s$ is not very close to
$M_H$.

\begin{figure}[htb]
\begin{center}
\begin{tabular}{c}
\epsfxsize=9truecm
\epsffile{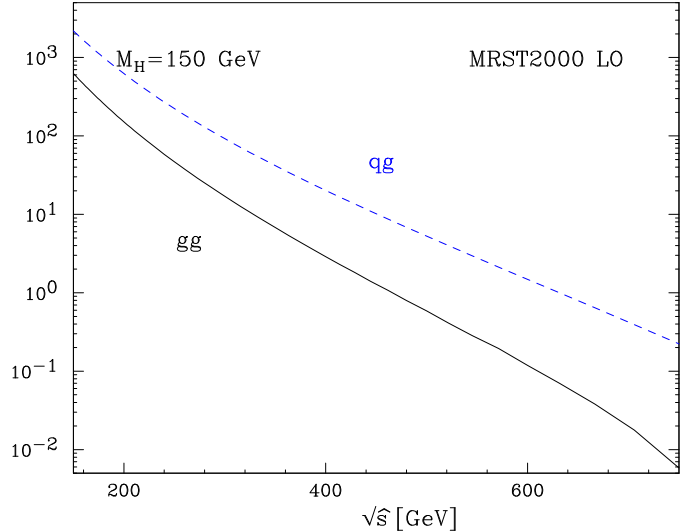}\\
\end{tabular}
\end{center}
\caption{\label{fig:l}{\em Parton ($gg$ and $qg+g\bar q$) luminosities 
${\cal L}(\tau,\mu_F^2)$ with $\mu_F=M_H=150$~GeV at the Tevatron Run II
as a function of the partonic centre-of-mass energy $\sqrt{{\hat s}}$ 
$({\hat s}= \tau s)$.
We use the LO set of parton distributions in Ref.~\cite{mrst2000}.}}
\end{figure}

At fixed $M_H$, the quantitative reliability of the large-$z$ approximation 
depends on the value of $\sqrt s$ and on the actual value of the parton
luminosities ${\cal L}_{ab/h_1h_2}(\tau,\mu_F^2)$ in Eq.~(\ref{sighad}). 
In Fig.~\ref{fig:l} the $gg$ and $(qg+g\bar q)$ luminosities 
(with $\mu_F=M_H=150$~GeV)
at the Tevatron Run II are plotted as a function of
the partonic centre-of-mass energy $\sqrt{{\hat s}}$ ($\hat s=\tau s$). 
We see that the luminosities decrease by almost two orders
of magnitude when $\sqrt{{\hat s}}$ increases from
$M_H=150$ GeV (hadronic threshold) to $300$ GeV.
A similar effect occurs at the LHC (see Fig.~1 in Ref.~\cite{moriond}).
We thus expect that the large-$z$ expansion of the coefficient function 
$G_{ab}(z)$ reliably approximates the size of the higher-order QCD corrections 
to Higgs boson production at the Tevatron and the LHC. At NLO, this was
explicitly checked in Ref.~\cite{Kramer:1998iq}, by comparison with the
complete expression of $G_{ab}^{(1)}(z)$. 

The authors of Ref.~\cite{Kramer:1998iq} also pointed out that at NLO the
numerical effect of the logarithmic term $\ln (1-z)$ of collinear origin
is not small. Following this observation, in Ref.~\cite{Catani:2001ic}
we introduced the soft-virtual-collinear (SVC) approximation of the coefficient
function $G_{ab}(z)$. The SVC approximation is defined by including
the leading $\ln^k(1-z)$ contribution from the collinear region
in the $gg$ channel:
\begin{equation}
\label{gg1SVC}
 G_{gg}^{(1){\rm SVC}}(z;M_H^2/\mu^2_R;M_H^2/\mu^2_F) =
G_{gg}^{(1){\rm SV}}(z;M_H^2/\mu^2_R;M_H^2/\mu^2_F) -12\, 
\ln(1-z) \;\;,
\end{equation}
\begin{equation}
\label{gg2SVC}
G_{gg}^{(2){\rm SVC}}(z;M_H^2/\mu^2_R;M_H^2/\mu^2_F) =
G_{gg}^{(2){\rm SV}}(z;M_H^2/\mu^2_R;M_H^2/\mu^2_F) - 72\, \ln^3(1-z) \;\;.
\end{equation}
Since the term $\ln^k(1-z)$ added to the SV expressions is that with the 
highest power $k$ at each perturbative order ($k=1$ and $k=3$ at LO and NLO,
respectively), the SVC approximation consistently
includes the next-to-dominant contribution to $G_{gg}^{(2)}$ in the limit $z \to 1$.
The comparison between the SV and the SVC approximations can thus be used 
\cite{Catani:2001ic} to gauge the quantitative accuracy of approximating
$G_{ab}(z)$ by its large-$z$ limit.
In the next section we study the impact of the SV and SVC 
approximations at the Tevatron Run II.

Two different large-$z$ approximations, named `soft' and `soft+sl' were
also considered in the numerical study of Ref.~\cite{Harlander:2001is}.
The `soft' approximation of Ref.~\cite{Harlander:2001is} regards the whole
partonic cross section ${\hat \sigma}_{ab}$ in Eq.~(\ref{sigmapart}),
while our SV approximation refers only to the hard coefficient function $G_{ab}(z)$.
In other words, we perform the soft approximation on the
phase-space integral of the matrix elements squared, while the {\em kinematical}
flux factor $1/{\hat s}$ in ${\hat \sigma}_{ab}$ is kept fixed and
not expanded around ${\hat s} = s$. This means that we expand $G_{ab}(z)$ 
around $z=1$, whereas the authors of Ref.~\cite{Harlander:2001is}
consider the expansion of ${\widetilde G}_{ab}(z)= z G_{ab}(z)$. 
Owing to the identity
\begin{equation}
\label{id}
z\, {\cal D}_i={\cal D}_i-\ln^i(1-z) \;,
\end{equation}
the two expansions differ by subdominant $\ln^i(1-z)$ contributions
of kinematical origin (see also Sect.~3 in Ref.~\cite{Catani:2001ic}).
Moreover, the `soft+sl' approximation of ${\widetilde G}_{gg}^{(2)}$ 
\cite{Harlander:2001is}
includes also additional subleading terms, proportional to $\ln^2 (1-z)$ and 
$\ln (1-z)$, whose coefficients (unlike those in Eqs.~(\ref{gg2SV}) and 
(\ref{gg2SVC})) are not (exactly) predictable at present \cite{Kramer:1998iq}.
In summary, we note that there is no one-to-one correspondence between 
`soft' (or `soft+sl')
in Ref.~\cite{Harlander:2001is} and SV (or SVC) in our definition, 
the difference being due to logarithmic contributions that, formally,
are consistently subdominant when $z \to 1$.

Although, from a formal viewpoint, it is legitimate to perform the large-$z$
expansion of either $G_{ab}(z)$ or ${\widetilde G}_{ab}(z)$, the two expansions
can quantitatively differ when applied to the evaluation of the hadronic cross
section in Eq.~(\ref{had}). In fact, when using ${\widetilde G}_{ab}(z)$, the
analogue of Eq.~(\ref{sighad}) is
\begin{equation}
\label{sighadcal}
\sigma(s,M_H^2)=\sigma_0\,
\sum_{a,b}\int_{\tau_H}^1 \f{d\tau}{\tau}\, 
{\cal \widetilde L}_{ab/h_1h_2}(\tau,\mu_F^2)\, 
{\widetilde G}_{ab}(\tau_H/\tau) \;.
\end{equation}
Comparing Eqs.~(\ref{sighad}) and (\ref{sighadcal}), we see that the coefficient
function ${\widetilde G}_{ab}(z=\tau_H/\tau)$ is convoluted with the
momentum-fraction luminosity 
${\cal \widetilde L}(\tau,\mu_F^2) = \tau {\cal L}(\tau,\mu_F^2)$.
Owing to the relative rescaling factor $\tau= {\hat s}/s$,
${\cal \widetilde L}(\tau,\mu_F^2)$ is much less steep than the luminosity
${\cal L}(\tau,\mu_F^2)$ (see Fig.~\ref{fig:l}) that enters Eq.~(\ref{sighad}).
Therefore, in the numerical evaluation of the Higgs boson cross section, we
expect and 
anticipate (see the comment at the end of Sect.~\ref{pheno}) that the large-$z$
expansion of ${\widetilde G}_{ab}(z)$ \cite{Kramer:1998iq,Harlander:2001is}
converges more slowly than that of 
$G_{ab}(z)$. The slowing down is ultimately caused by a too extreme kinematics
approximation of the partonic cross section in Eq.~(\ref{sigmapart}): the flux
factor $1/{\hat s}$ has been replaced by $1/M_H^2$.

\section{Inclusive production 
at the Tevatron Run II}
\label{pheno}

In this section we
study the higher-order QCD
corrections to the inclusive production of the SM Higgs boson at the Tevatron 
Run II, i.e. proton--antiproton collisions at
$\sqrt{s}=2$~TeV. 
We recall that we include the exact dependence on $M_{\rm top}$ in the Born-level 
cross section $\sigma_0$ (see Eq.~(\ref{borncs})), while the coefficient 
function $G_{ab}(z)$ is evaluated in the large-$M_{\rm top}$ limit. At NLO 
\cite{Spira:1995rr, Kramer:1998iq}
this is a very good numerical approximation when $M_H \leq 2M_{\rm top}$.

 Unless otherwise stated, cross sections are computed
 using the MRST2000 \cite{mrst2000} sets of
 parton distributions with densities and coupling constant evaluated at each
 corresponding order, i.e. using LO distributions 
and 1-loop $\as$ for the LO
 cross section, and so forth. The corresponding values of
 $\Lambda^{(5)}_{QCD}$ ($\as(M_Z)$)  are  $0.132$ (0.1253), $0.22$ (0.1175)
   and $0.187$ (0.1161)~GeV, at 1-loop, 2-loop and 3-loop order,
   respectively.
 In the  NNLO case  we  use the `central' set of
 MRST2000, obtained from a global fit of data (deep inelastic scattering,
 Drell--Yan production and jet $E_T$ distribution) by using the approximate
 NNLO evolution kernels presented in Ref.~\cite{vnvogt}. 
The result we refer to as NNLO-SV(SVC) corresponds to the sum of the
LO and exact NLO (including the $qg$ and $q\bar{q}$ channels)
contributions plus the SV(SVC) corrections at NNLO,
given in Eq.~(\ref{gg2SV}) (Eq.~(\ref{gg2SVC})).
\begin{figure}[htb]
\begin{center}
\begin{tabular}{c}
\epsfxsize=9truecm
\epsffile{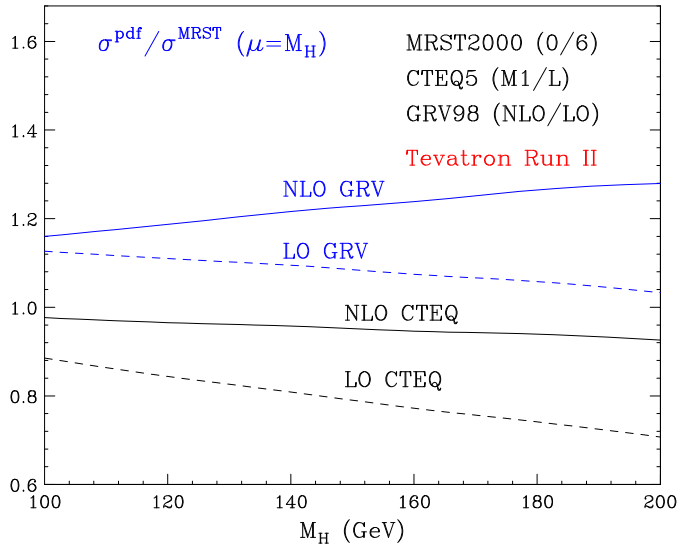}\\
\end{tabular}
\end{center}
\caption{\label{fig:pdf}{\em Relative difference of LO and NLO results
computed with GRV98 and CTEQ5 distributions with respect to MRST2000
distribution. }}
\end{figure}

In Fig.~\ref{fig:pdf} we show the dependence of the LO and NLO cross sections
on the choice of parton distributions.
The results obtained by using both the GRV98 \cite{Gluck:1998xa} and 
CTEQ5 \cite{cteq5} sets 
differ substantially from the MRST2000 results.
In the case of the GRV98 sets, this difference is not unexpected, since the
value of 
$\as(M_Z)$ and
the gluon distribution are quite different from those of 
MRST2000 and CTEQ5.
We see that, as $M_H$ increases from $100$ to
$200$~GeV, the LO result obtained by using the CTEQ5 set is between
$10$ and $30\%$ lower than the one obtained by MRST2000.
At NLO the difference is smaller, and it 
is compatible with the $\pm 10\%$ uncertainty recommended
by the CTEQ collaboration \cite{Huston:1998jj} on $gg$ and $qg$
luminosities in the $x$ region that controls Higgs boson production at the Tevatron.
Although the LO differences are physically less meaningful than the NLO
differences, the size of the former has to be kept in mind when quoting
QCD predictions (based on either analytic calculations or Monte Carlo event
generators) that directly or indirectly (e.g. in the case of $K$-factors)
use LO parton densities.

\begin{figure}[htb]
\begin{center}
\begin{tabular}{c}
\epsfxsize=15truecm
\epsffile{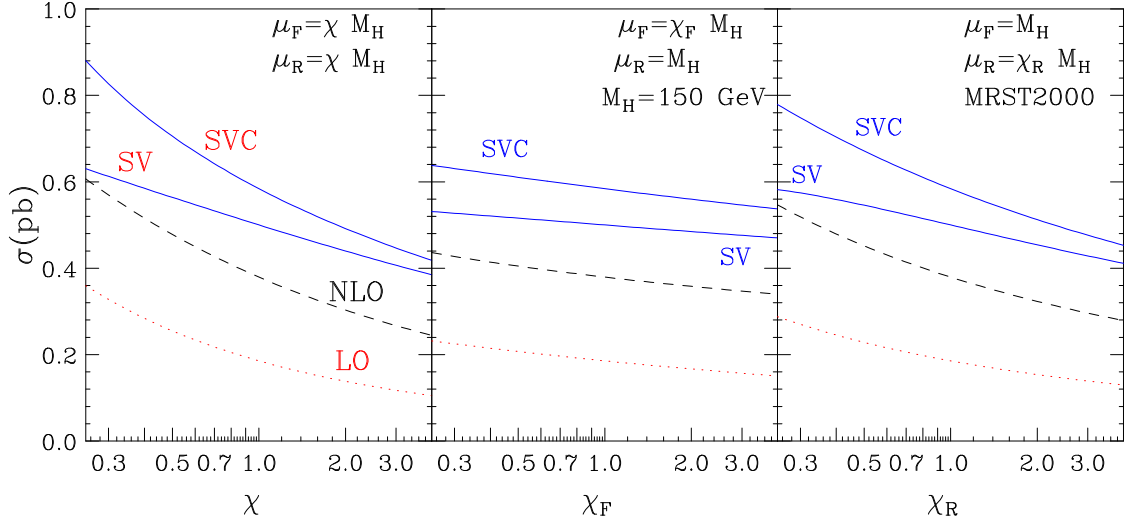}\\
\end{tabular}
\end{center}
\caption{\label{fig:scale}{\em Scale dependence of the Higgs production cross section for
    $M_H=150$~GeV at LO, NLO, NNLO-SV and NNLO-SVC. }}
\end{figure}

Figure~\ref{fig:scale} shows
the scale dependence of the cross section for the production of a Higgs boson
with $M_H=150$~GeV. The scale dependence is analysed by varying the factorization
 and renormalization scales by a factor of 4 up and down from the central value $M_H$. 
The plot on the left corresponds to the simultaneous variation of both scales,
 $\mu_F=\mu_R=\chi \, M_H$,
whereas the plots in the centre and on the right respectively correspond to 
the results of the independent variation of the factorization and 
renormalization scales, keeping the other fixed at the central value.

As expected from the QCD running of $\as$, the cross sections typically 
decrease when $\mu_R$ increases around the characteristic hard scale $M_H$.
A similar behaviour\footnote{At the LHC, the $\mu_F$ dependence is opposite,
as shown in Fig.~1 of Ref.~\cite{Catani:2001ic}.}
is observed when $\mu_F$ varies, since the cross sections
are mainly sensitive to partons with momentum fraction $x \sim 0.05$--$0.1$, and in
this $x$ range scaling violation of the parton densities is (slightly) negative.
The largest variations in the 
cross section calculation are thus obtained by simultaneously varying $\mu_R$
and $\mu_F$.
The scale dependence is 
mostly driven by the renormalization scale, because the lowest-order 
contribution  to the process is proportional to $\as^2$, 
a (relatively) high power of $\as$. 

In summary, Fig.~\ref{fig:scale} shows that
the scale dependence is reduced when higher-order corrections are included. 
Varying the scales in the range $0.5 \leq \chi_R, \chi_F \leq 2$,
the reduction is 
from about $\pm 20 \%$ at full NLO to about $\pm 10 \%$ and $\pm 15 \%$
at NNLO-SV and NNLO-SVC, respectively.
The increase in the scale dependence when going from NNLO-SV to NNLO-SVC
is due to the fact that the contribution of
the dominant collinear terms included in the SVC
approximation (see Eqs.~(\ref{gg1SVC}) and (\ref{gg2SVC})) is not small and 
scale-independent, so it cannot be compensated by scale variations.

\begin{figure}[htb]
\begin{center}
\begin{tabular}{c}
\epsfxsize=12truecm
\epsffile{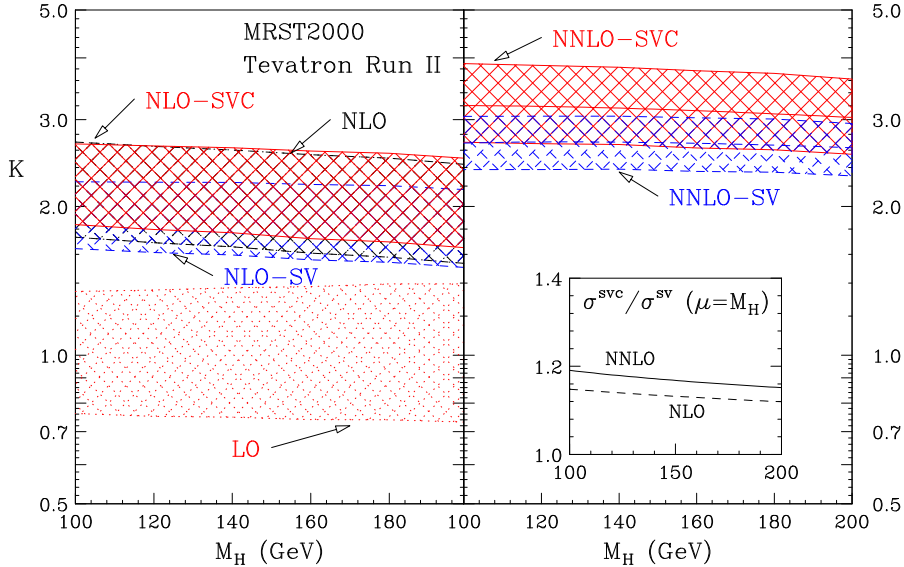}\\
\end{tabular}
\end{center}
\caption{\label{fig:k1}{\em  $K$-factors for Higgs production for the full NLO result
and the NLO-SV, NLO-SVC, NNLO-SV and NNLO-SVC approximations.}}
\end{figure}

In Fig.~\ref{fig:k1} we study the
$K$-factors, defined as the ratio of the cross section evaluated at
each corresponding order over the LO result.
Since the LO result sizeably depends on the choice of parton distributions, 
the $K$-factor should be interpreted
with care. For instance, for $M_H=180$~GeV, the NLO $K$-factor computed by
using MRST distributions is about $20\%$ smaller than the one obtained by
using CTEQ distributions.
  
In Fig.~\ref{fig:k1} the bands account for the 
`theoretical uncertainty' due to 
the scale dependence, quantified by using the  minimum and maximum values of 
the cross sections when the scales $\mu_R$ and $\mu_F$
are varied (simultaneously and independently, as in Fig.~\ref{fig:scale})
 in the range $0.5 \le  \chi_R, \chi_F\le 2$. The LO result that 
 normalizes the $K$-factors is computed at the default
scale $M_H$ in all cases.

The plot on the left-hand side of Fig.~\ref{fig:k1} shows
the scale uncertainty at LO and compares the full NLO result with the NLO-SV and
 NLO-SVC approximations. When $M_H$ is in the range $140$--$180$~GeV,
the NLO-SV approximation  
 tends to underestimate the full NLO result by about $7$--$8\%$, whereas the 
 NLO-SVC approximation overestimates it by about $4$--$5\%$,
showing the 
effect 
of the term $\ln(1-z)$ added in the SVC approximation.
We find that the SVC result provides an excellent approximation of the full $gg$ contribution at NLO,
the difference being less than 2
per mille.
The contribution of the $qg$ channel tends to lower the SVC result
by $\ltap 5\%$. The effect of the $qg$ channel is mainly due to the
logarithmically-enhanced behaviour
$G^{(1)}_{qg}(z) \simeq \frac{4}{3} \ln (1-z)$ of the corresponding coefficient
function at large $z$.
The improved reliability of the SV and SVC approximations with respect to the
LHC \cite{Catani:2001ic} is not unexpected since
at the Tevatron we are closer to threshold.

The right-hand side of Fig.~\ref{fig:k1} shows the SV and SVC results at NNLO. Again, 
the SVC band
sits higher
than the SV one and, as shown in the inset plot, 
the ratio of the corresponding cross sections 
increases with respect to NLO.
The contribution from non-leading terms $\ln^k(1-z)$, with $k<3$
(which are not under control within the SVC approximation), is not included.
We have tried to add
a term $\ln^2(1-z)$ with a coefficient as large as the one of the
$\ln^3(1-z)$ contribution, and we have found a small (${\cal O}(5\%)$) modification.
Therefore, we expect
the effect of these non-leading logarithmic terms to
be numerically less important.
As pointed out in Ref.~\cite{Catani:2001ic}, at NNLO 
there is a (still unknown) leading collinear contribution proportional to
$\ln^3(1-z)$ also in the $qg$ channel.
Owing to the size of the contribution of the $qg$ channel at NLO,
the quantitative effect of
this NNLO term 
can be of the same order 
as the corresponding one included through the SVC approximation
in the $gg$ channel.

We recall that the NNLO results on the right-hand side of Fig.~\ref{fig:k1}
are obtained by using the (approximated) NNLO parton distributions of the
MRST2000 set. Using the NLO parton distributions, the $K$-factors would be
smaller by about 5 to $8\%$ as $M_H$ increases from 120 to 180~GeV.

In summary, considering the results obtained at NLO,
we expect the 
full NNLO $K$-factor
to be between the SV and SVC bands. 
For reference, we give in Table 1 the central values of the cross section
in the range $M_H=140$--$180$~GeV. 
In particular, for a light Higgs boson ($M_H \ltap 200$~GeV),
this 
corresponds to an increase of about
$50\%$ with respect to the full NLO result, i.e. a factor of $\sim 3$
with respect to the LO result.
Taking into account that the NLO result increases the
LO cross section by a factor of about 2, our result shows
that the convergence of the
perturbative series is poorer at the Tevatron than at the LHC 
\cite{Catani:2001ic}. This also implies that QCD contributions
beyond NNLO can still be significant at the Tevatron.

Our conclusion on the possible relevance of higher-order contributions is not 
in contradiction with the improved scale dependence of the NNLO calculation.
As is well known, the customary procedure of varying
the scales to estimate the theoretical uncertainty due to missing higher-order
terms can only give a lower limit on the `true' uncertainty. 
This is well demonstrated by the plot on the left-hand side of
Fig.~\ref{fig:k1}, which
shows no overlap between the LO and NLO bands. Since the NLO and NNLO bands
still tend to show no (or a marginal) overlap, their size cannot yet
be regarded as a 
reliable estimate of the theoretical uncertainty.

\begin{table}[t]
\begin{center}
\begin{tabular}{|c|c|c|c|c|c|c|}
   \hline $M_H$(GeV) & LO & NLO & NLO-SV & NLO-SVC & NNLO-SV & NNLO-SVC\\
   \hline $140$ & 0.2282 & 0.4715 & 0.4338 & 0.4922 & 0.6163 & 0.7222\\
\hline $150$ & 0.1856 & 0.3794 & 0.3507 & 0.3969 & 0.5002 & 0.5841\\
\hline $160$ & 0.1523 & 0.3080 & 0.2859 & 0.3229 & 0.4092 & 0.4763\\
\hline $170$ & 0.1260 & 0.2519 & 0.2349 & 0.2646 & 0.3369 & 0.3912\\
\hline $180$ & 0.1050 & 0.2075 & 0.1943 & 0.2184 & 0.2795 & 0.3235\\
\hline
\end{tabular}
\caption{\label{table} \em Cross sections in $pb$ as a function of $M_H$.
The calculation is performed by setting $\mu_R=\mu_F=M_H$.}
\end{center}
\end{table}

A comment on the numerical results presented in Ref.~\cite{Harlander:2001is} is in order.
The authors of Ref.~\cite{Harlander:2001is} use different parton distributions 
and, as explained in Sect.~\ref{sec:theo}, there is no one-to-one 
correspondence between
their `soft' (or `soft+sl') approximation and our SV (or SVC) approximation. 
Therefore, those results cannot directly be compared with ours
(see also Ref.~\cite{Harlander:2001eb}). Certainly, they 
show a slower numerical convergence of the large-$z$ expansion used in 
Ref.~\cite{Harlander:2001is}.
As discussed at the end of Sect.~\ref{sec:theo}, this is not unexpected because of
significant subdominant effects of kinematical origin.

\section{Vetoed cross section}
\label{sec:veto}

Direct Higgs production followed by
the decay $H\to WW\to l^+l^-\nu{\bar \nu}$ is an important channel to 
discover the SM Higgs boson in the intermediate-mass range 
140~GeV~$\ltap M_H \ltap$~190~GeV both at the Tevatron and at the
LHC. Nevertheless, as recalled in Sect.~\ref{sec:intro}, several experimental
cuts have to be applied to discriminate the signal over the background.
An important selection to enhance the statistical significance is a veto
on the high transverse-momentum jets that accompany the production of the
Higgs boson. Events with high transverse-momentum jets are excluded from the
analysis \cite{Carena:2000yx,atlascms,Han:1999ma}.

In this section we study the effect of a jet veto on inclusive Higgs
production.
The events that pass the veto selection are those with
$p_T^{\rm jet} < p_T^{\rm veto}$, where $p_T^{\rm jet}$ is the transverse
momentum of any final-state jets.
Jets are defined by a cone algorithm.
In the perturbative calculation, jets are represented by a parton or a set
of
partons. The Higgs boson can be accompanied by one final-state parton
at NLO, and by one or two final-state partons at NNLO.
In the case of a single final-state parton with
transverse momentum ${\bf p}_{1 \,T}$,
the vetoed cross section is computed by imposing
$| {\bf p}_{1 \,T}| < p_T^{\rm veto}$ (i.e. we veto events with $| {\bf
p}_{1
\,T}| > p_T^{\rm veto}$).
When there are two final-state partons with transverse momenta
${\bf p}_{1 \,T}$ and ${\bf p}_{2 \,T}$, we consider their angular
distance
$R_{12}^2= (\eta_1 - \eta_2)^2 + (\phi_1 - \phi_2)^2$ in the
pseudorapidity--azimuth plane: if $R_{12} > R$, we impose the constraints
$| {\bf p}_{1 \,T}|, | {\bf p}_{2 \,T}| < p_T^{\rm veto}$ (i.e. we veto
events
with $| {\bf p}_{1 \,T}| > p_T^{\rm veto}$ or
$| {\bf p}_{2 \,T}| > p_T^{\rm veto}$);
if $R_{12} < R$, we combine the two partons in a single jet and we impose
$| {\bf p}_{1 \,T} + {\bf p}_{2 \,T}| < p_T^{\rm veto}$
(i.e. we veto events with
$| {\bf p}_{1 \,T} + {\bf p}_{2 \,T}| > p_T^{\rm veto}$).
In all the numerical
results presented in this section, the cone size $R$ of the jets is fixed
at the
value $R=0.4$.

As in the case of the inclusive cross section, we evaluate the 
jet-vetoed cross section by using 
the large-$M_{\rm top}$ limit. Studies on Higgs+jets production show that
the QCD matrix elements in the large-$M_{\rm top}$ limit 
\cite{Schmidt:1997wr,Dawson:1992au} approximate very well those with the exact
$M_{\rm top}$ dependence \cite{Ellis:1988xu,DelDuca:2001eu}, provided the 
jet transverse momenta and, hence, $p_T^{\rm veto}$ remain small with respect to
$M_{\rm top}$.

The vetoed cross section $\sigma^{\rm veto}(s,M_H^2;p_T^{\rm veto},R)$ can be
written as 
\begin{equation}
\label{sigmaveto}
\sigma^{\rm veto}(s,M_H^2;p_T^{\rm veto},R) = \sigma(s,M_H^2) 
- \Delta \sigma(s,M_H^2;p_T^{\rm veto},R) \;\;,
\end{equation}
where $\sigma(s,M_H^2)$ is the inclusive hadronic cross section in 
Eq.~(\ref{had}), and $\Delta \sigma$ is the `loss' in cross section
due to the jet-veto procedure. 
The vetoed cross section 
is computable by a factorization formula analogous to Eq.~(\ref{had}),
apart from the replacement of the coefficient function $G_{ab}(z)$ by
the vetoed coefficient function $G_{ab}^{\rm veto}(z;\pitcut,R)$.
By analogy with Eq.~(\ref{sigmaveto}), we can write the vetoed coefficient 
function as
\begin{equation}
\label{coeffveto}
G_{ab}^{\rm veto}(z;\pitcut,R) = G_{ab}(z) - \Delta G_{ab}(z;\pitcut,R) \;\;,
\end{equation}
where, to simplify the notation (see Eq.~(\ref{deltag1})),
the dependence on $p_T^{\rm veto}$ is parametrized by the dimensionless variable
$\pitcut$:
\begin{equation}
\label{ptoverm}
\pitcut(z,p_T^{\rm veto}/M_H) \equiv \frac{2p_T^{\rm veto} {\sqrt z}}{(1-z)M_H}
\;\;.
\end{equation}

At LO, the vetoed cross section is equal to the inclusive one, so
the subtracted coefficient function $\Delta G_{ab}$ vanishes.
At higher perturbative orders, $\Delta G_{ab}$ is computable
according to the power-series expansion
\begin{align}
\label{dgexpansion}
\Delta G_{ab}(z;\pitcut,R;\as(\mu_R^2), M_H^2/\mu_R^2;M_H^2/\mu_F^2) 
&= \f{\as^3(\mu_R^2)}{\pi} \;
\Delta G_{ab}^{(1)}\left(z;\pitcut\right) \;
\Theta\!\left(1-\pitcut(z,p_T^{\rm veto}/M_H)\right) \nn \\
&+ \f{\as^4(\mu_R^2)}{\pi^2} \;
\Delta G_{ab}^{(2)}\left(z;\pitcut,R;\frac{M_H^2}{\mu_R^2};\frac{M_H^2}{\mu_F^2}\right) 
+ {\cal O}(\as^5) \;.
\end{align}
The NLO contribution $\Delta G_{ab}^{(1)}$ is scale-independent and
$R$-independent, but it depends on $p_T^{\rm veto}$. We have
explicitly evaluated it in the large-$M_{\rm top}$ limit, and we find:
\begin{align}
\Delta G_{gg}^{(1)}(z;\pitcut)&= {\hat P}_{gg}(z)
\;\ln\f{1+\sqr}{1-\sqr} - \f{11}{2} \; \f{(1-z)^3}{z}
\left( 1 - \f{\pitcut^2}{22} \right) \sqr \;,\nn \\
\label{deltag1}
\Delta G_{gq}^{(1)}(z;\pitcut)&= \f{1}{2}{\hat P}_{gq}(z)
\;\ln\f{1+\sqr}{1-\sqr} - \f{(1-z)^2}{z}\sqr \;, \\
\Delta G_{q\bar{q}}^{(1)}(z;\pitcut)&=
\f{32}{27}\;\f{(1-z)^3}{z} \left( 1- \frac{\pitcut^2}{4} \right) \sqr \;, \nn
\end{align}
where
\begin{align}
{\hat P}_{gq}(z)&=\f{4}{3} \;\f{1+(1-z)^2}{z} \;,\nn\\
{\hat P}_{gg}(z)&=6\left[ \f{1-z}{z}+\f{z}{1-z}+z(1-z)\right] \, .
\end{align}
The evaluation of the NNLO contribution $\Delta G_{ab}^{(2)}$ cannot be easily
performed in analytic form, since the calculation depends on the details of
the jet algorithm. We compute $\Delta G_{ab}^{(2)}$ numerically by using
the program of Ref.~\cite{deFlorian:1999zd}. The program, 
which implements the matrix elements (in the large-$M_{\rm top}$ limit) 
of Refs.~\cite{Schmidt:1997wr,Dawson:1992au} by using the subtraction-method 
procedure of Ref.~\cite{Frixione:1996ms},
computes the QCD corrections to Higgs+jet(s) production up to order 
$\as^4$. For the purpose of evaluating the contribution $\Delta \sigma$ to
Eq.~(\ref{sigmaveto}), we use the program to compute the Higgs+jet(s)
cross section when the transverse momentum of the highest-$p_T$ jet is larger 
than $p_T^{\rm veto}$.

In the following we present both NLO and NNLO numerical results for the vetoed
cross section $\sigma^{\rm veto}$ in Eq.~(\ref{sigmaveto}). 
The results are obtained by using the
parton distributions of the MRST2000 set, as explained in Sect.~\ref{pheno}.
The NLO calculation is exact: apart from using the 
large-$M_{\rm top}$ limit, we do not perform any further approximations.
At the NNLO, the contribution $\Delta \sigma$ to Eq.~(\ref{sigmaveto})
is again evaluated exactly, 
while to evaluate the contribution of the inclusive
cross section we rely on our approximate estimate in Sect.~\ref{pheno}
(see the corresponding LHC results in Sect.~4 of Ref.~\cite{Catani:2001ic})
and, in particular, we use the NNLO-SVC result.
Therefore, once the full NNLO result for the inclusive cross section is 
available, it can straightforwardly be used to `correct' our NNLO estimate for 
the vetoed cross section. As stated above, in our numerical calculations
we fixed the cone size of the jets to the value $R=0.4$. The $R$-dependence
of the perturbative calculation first appears at the NNLO. In particular, the
NNLO vetoed cross section decreases by increasing $R$.

Note that the numerical program of Ref.~\cite{deFlorian:1999zd} is a 
Monte Carlo code that evaluates Higgs+jet(s) production at the fully
exclusive level.
Therefore, it can be used through the subtraction procedure of 
Eq.~(\ref{sigmaveto}) to compute vetoed cross sections also when 
additional kinematical cuts (e.g. cuts on the rapidities of the jets, or
asymmetric cuts on the jet transverse momenta) or different jet definitions are
considered.

\begin{figure}[ht]
\begin{center}
\begin{tabular}{c}
\epsfxsize=11truecm
\epsffile{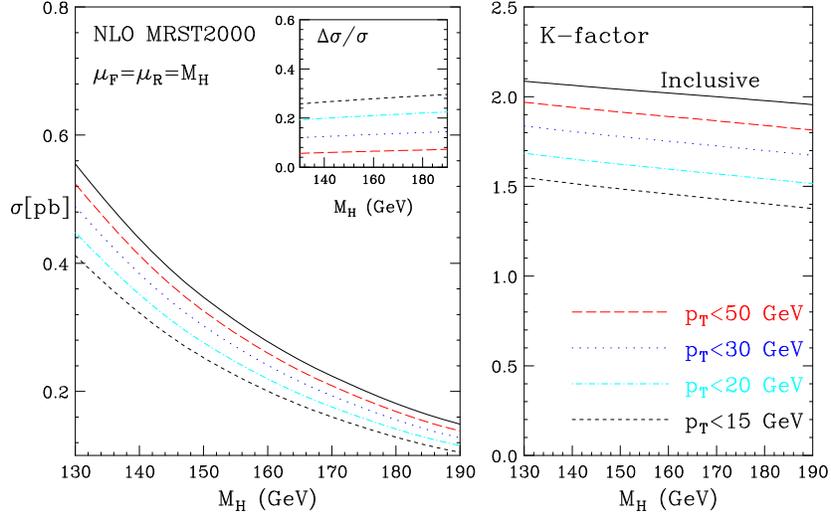}\\
\end{tabular}
\end{center}
\caption{\label{fig:nlotev}{\em Vetoed cross section
and $K$-factors: NLO results at the Tevatron Run II.}}
\end{figure}
We first present the vetoed cross section\footnote{The numerical program
of Ref.~\cite{deFlorian:1999zd} implements the large-$M_{\rm top}$ limit
strictly, i.e. also the Born-level contribution $\sigma_0$ in
Eq.~(\ref{borncs})
is evaluated in the limit $M_H/M_{\rm top} \to 0$. For simplicity, the
numerical
results of this section implement the same approximation. The
approximation used
in Sect.~\ref{pheno} can be recovered by simply rescaling the absolute
value of
the cross sections by the overall factor $\sigma_0/\sigma_0(M_H/M_{\rm
top}=0)$.
Of course, such a rescaling has no effects
on the ratios
$\Delta \sigma/\sigma$ and on the vetoed $K$-factors.}
at the Tevatron Run II.
In Fig.~\ref{fig:nlotev} we show the dependence of the NLO results 
on the Higgs mass for different values of 
 $p_T^{\rm veto}$ (15, 20, 30 and 50~GeV).
The vetoed cross sections $\sigma^{\rm veto}(s,M_H^2;p_T^{\rm veto},R)$
and the inclusive cross section $\sigma(s,M_H^2)$ are given in the plot 
on the left-hand side. The inset plot gives an idea of
the `loss' in cross section once the veto is applied,
by showing the ratio between the cross section difference $\Delta \sigma$ 
in Eq.~(\ref{sigmaveto}) and the inclusive cross section at the same
perturbative order.
As can be observed, for large values of the cut, say $p_T^{\rm veto}=50$~GeV,
less than 10\% of the inclusive cross section is vetoed. The veto effect
increases by decreasing $p_T^{\rm veto}$, but it is still smaller than 30\%
when $p_T^{\rm veto}=15$~GeV.
On the right-hand side of Fig.~\ref{fig:nlotev},
we show the corresponding $K$-factors, i.e. the vetoed
cross sections normalized to the LO result,
which is independent of the value of the cut.
\begin{figure}[htb]
\begin{center}
\begin{tabular}{c}
\epsfxsize=11truecm
\epsffile{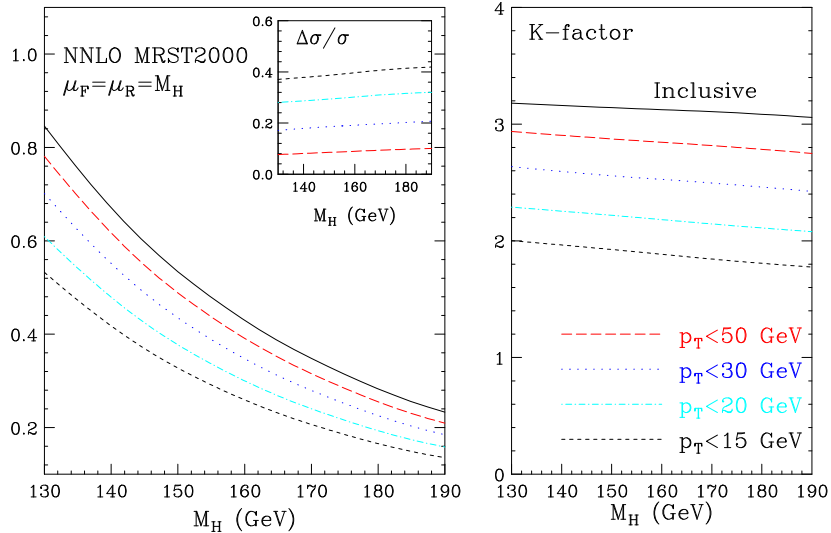}\\
\end{tabular}
\end{center}
\caption{\label{fig:nnlotev}{\em Vetoed cross section and $K$-factors:
NNLO results at the Tevatron Run II.}}
\end{figure}

Figure~\ref{fig:nnlotev} shows the analogous results at NNLO.
Note that, although $\Delta \sigma/\sigma$ is slightly larger than at NLO,
the vetoed cross sections and, thus,
the $K$-factors, are still larger than at NLO. This is mostly due
to the large increase of the inclusive cross section at NNLO,
as shown in Sect.~\ref{pheno}.

\begin{figure}[htb]
\begin{center}
\begin{tabular}{c}
\epsfxsize=10truecm
\epsffile{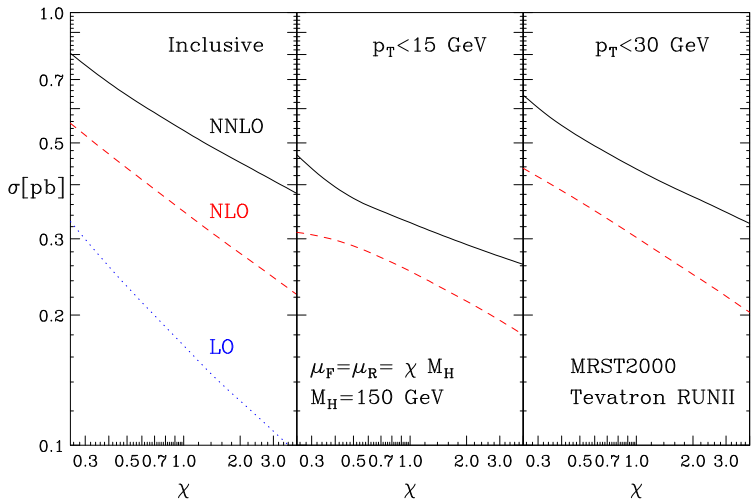}\\
\end{tabular}
\end{center}
\caption{\label{fig:scalecut}{\em Scale dependence
of the inclusive and vetoed cross sections at LO, NLO and NNLO-SVC at the 
Tevatron.}}
\end{figure}
All the results plotted in Figs.~\ref{fig:nlotev} and \ref{fig:nnlotev}
have been obtained by fixing the renormalization and factorization scales
at the default value $\mu_R=\mu_F=M_H$. In Fig.~\ref{fig:scalecut},
we study the scale dependence of the cross sections
(inclusive and with $p_T^{\rm veto}=15$ and $30$~GeV)
at LO, NLO and NNLO.
The renormalization and factorization scales are varied simultaneously by
a factor of 4 up and down with respect to $M_H$.
No significant differences are found when the scales
are varied separately. As can be observed, there is an improvement
in the stability of the result when higher order corrections are included.
This is particularly noticeable when going from LO to NLO, while
the comparison between NNLO and NLO looks rather similar
to the one observed in the inclusive case.

\begin{figure}[htb]
\begin{center}
\begin{tabular}{c}
\epsfxsize=8truecm
\epsffile{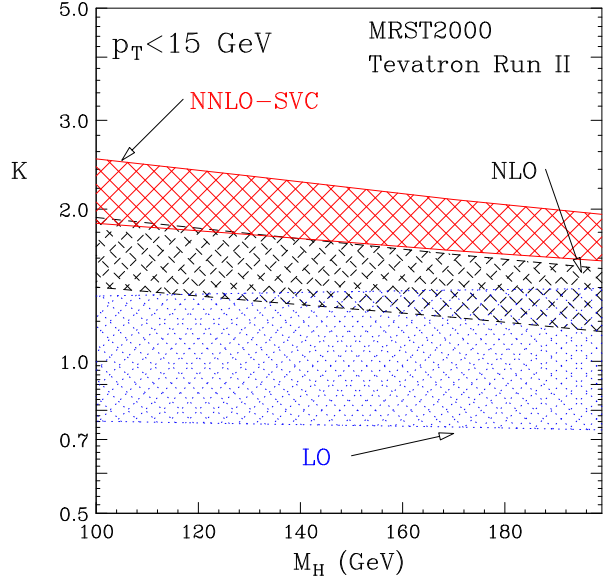}\\
\end{tabular}
\end{center}
\caption{\label{fig:kv}{\em  $K$-factors for Higgs production
at the Tevatron in the case of a veto of $p_T^{\rm veto}=15$~GeV: LO, NLO
and NNLO-SVC approximation.}}
\end{figure}
The scale-dependence effects on the perturbative $K$-factors can be appreciated
also from Fig.~\ref{fig:kv}, where the LO, NLO and NNLO-SVC bands
(computed as in Fig.~\ref{fig:k1}) are shown in
the case in which a veto of $p_T^{\rm veto}=15$~GeV is applied.
Comparing Fig.~\ref{fig:kv} with the inclusive case in Fig.~\ref{fig:k1},
we see that the effect of the veto is to partially reduce the
relative difference between the NLO and NNLO results; the increase of 
the corresponding $K$-factors
can be estimated to about $25\%$.

The results on the jet veto presented so far can be qualitatively explained by 
a simple physical picture. As shown in Sect.~\ref{pheno}, the effect of
the higher-order contributions to inclusive Higgs boson production
at the Tevatron is large. The dominant part of this effect
is due to soft and collinear radiation, whereas the accompanying hard
radiation has little effect. The characteristic scale of the highest
transverse momentum $p_T^{\rm max}$ of the accompanying jets is indeed 
$p_T^{\rm max}\sim \langle 1- z \rangle M_H$ (see e.g. the upper bound on
$p_T^{\rm veto}$ from the theta function in Eq.~(\ref{dgexpansion})), 
where the average value 
$\langle 1- z \rangle = \langle 1 - M_H^2/{\hat s}\rangle$ of the distance
from the partonic threshold is small. As a consequence the jet veto procedure
is weakly effective unless the value of $p_T^{\rm veto}$ is pretty small
(i.e. substantially smaller than $p_T^{\rm max}$).
Decreasing $p_T^{\rm veto}$,
the enhancement of the inclusive cross section due to soft radiation at higher
orders is reduced, and the jet veto procedure tends to
improve the convergence of the perturbative series (see e.g. Figs.~\ref{fig:k1}
and \ref{fig:kv}).
Note also that the characteristic scale $p_T^{\rm max}$
is a slightly increasing function of $M_H$, the linear increase with $M_H$
being partially compensated by the decrease of $\langle 1- z \rangle$.
Therefore, at fixed $p_T^{\rm veto}$ the vetoed $K$-factor decreases more than
the inclusive $K$-factor when $M_H$ increases
(see Fig.~\ref{fig:nnlotev}).

On the basis of this physical picture, we can easily anticipate
the qualitative effect of the jet veto at the LHC.
The overall features of the QCD corrections to
inclusive Higgs boson production at the LHC
\cite{Catani:2001ic} are the same as at the Tevatron. The main quantitative
differences arise from the fact that Higgs production at the LHC is less
close to threshold than at the Tevatron and, therefore, the accompanying jets
are less soft ($\langle 1- z \rangle$ is larger) at the LHC than at the 
Tevatron. At fixed $p_T^{\rm veto}$ the effect of the jet veto is thus
stronger at the LHC.
\begin{figure}[htb]
\begin{center}
\begin{tabular}{c}
\epsfxsize=11truecm
\epsffile{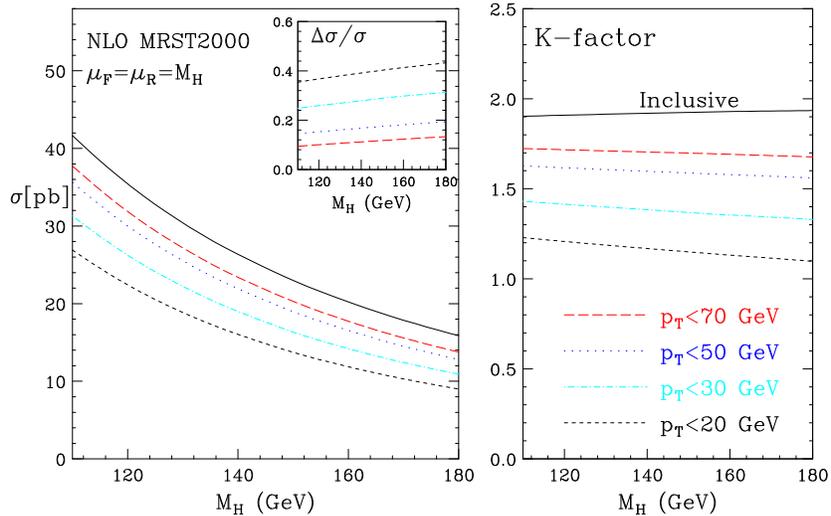}\\
\end{tabular}
\end{center}
\caption{\label{fig:nlolhc}{\em Vetoed cross sections and
$K$-factors at NLO at the LHC. }}
\end{figure}
\begin{figure}[htb]
\begin{center}
\begin{tabular}{c}
\epsfxsize=11truecm
\epsffile{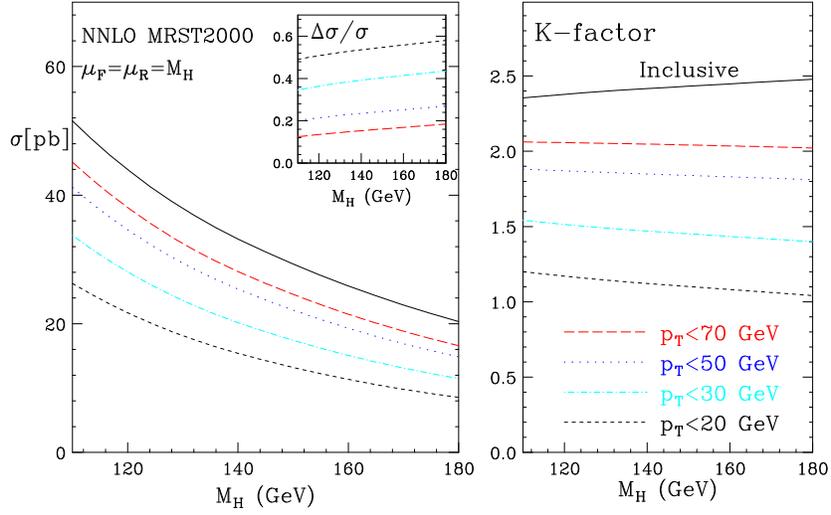}\\
\end{tabular}
\end{center}
\caption{\label{fig:nnlolhc}{\em Vetoed cross sections and
$K$-factors at NNLO at the LHC.  }}
\end{figure}
 
\begin{figure}[htb]
\begin{center}
\begin{tabular}{c}
\epsfxsize=8truecm
\epsffile{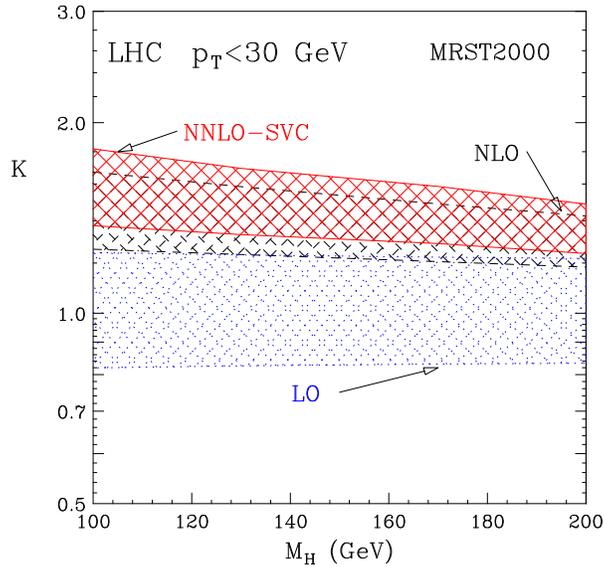}\\
\end{tabular}
\end{center}
\caption{\label{fig:kvlhc}{\em  The same as in Fig.~\ref{fig:kv}, but at the LHC
and with $p_T^{cut}=30$~GeV.}}
\end{figure}

The results for the vetoed cross sections at the LHC are presented in
Figs.~\ref{fig:nlolhc} and \ref{fig:nnlolhc} for 
$p_T^{\rm veto}=20$, 30, 50 and 70~GeV.
At fixed value of the cut, the impact of the jet veto, 
both in the `loss' of cross section and in the reduction of the $K$-factors,
is larger at the LHC than at the Tevatron Run II. This effect can also be 
appreciated by comparing Fig.~\ref{fig:kvlhc} and Fig.~\ref{fig:kv}.
At the LHC, the value of $p_T^{\rm veto}=30$~GeV is already sufficient to reduce
the difference between the NNLO and NLO results
to less than $10\%$. 

\begin{figure}[htb]
\begin{center}
\begin{tabular}{c}
\epsfxsize=8truecm
\epsffile{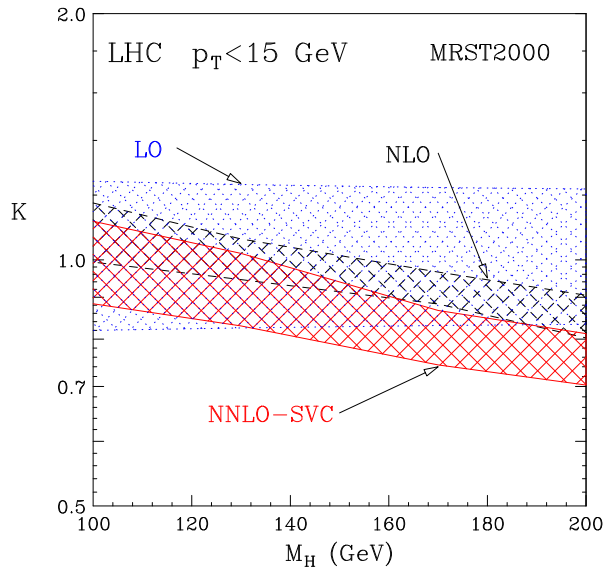}\\
\end{tabular}
\end{center}
\caption{\label{fig:kvlhc15}{\em  The same as in Fig.~\ref{fig:kv}, but at the LHC
and with $p_T^{cut}=15$~GeV.}}
\end{figure}

Note that when $p_T^{\rm veto}$ is much smaller than the characteristic scale
$p_T^{\rm max}\sim \langle 1- z \rangle M_H$, the coefficients
of the perturbative expansion
of the vetoed cross section contain large logarithmically-enhanced
contributions. For instance, from Eq.~(\ref{deltag1}) we have
\begin{equation}
\Delta G_{gg}^{(1)}(z;\pitcut) \sim 2 {\hat P}_{gg}(z)
\;\ln\f{(1-z)M_H}{p_T^{\rm veto}} \;.
\end{equation}
The presence of these contributions can spoil the quantitative convergence
of the fixed-order expansion in $\as$. Since $\langle 1- z \rangle M_H$ is
larger at the LHC than at the Tevatron, the value of $p_T^{\rm veto}$ at which
these effects become visible is larger at the LHC. The perturbative $K$-factors
shown in Fig.~\ref{fig:kvlhc15} suggest that at the LHC the vetoed cross 
section is sensitive to these large logarithmic terms already when
$p_T^{\rm veto}=15$~GeV. Indeed, the scale-dependence band is larger at NNLO
than at NLO.
At such small values of $p_T^{\rm veto}$, perturbative contributions beyond the
NNLO can still be significant.

\section{Conclusions}
\label{sec:conc}

In this paper we have presented our study of QCD radiative corrections to direct Higgs boson
production at hadron colliders and the impact of jet veto.

Using the theoretical results of Refs.~\cite{Catani:2001ic,Harlander:2001is},
in Sect.~\ref{pheno} we have first
shown that the NNLO-SV and NNLO-SVC approximations
are expected to be a good estimate of the full NNLO contributions
to the inclusive cross section at the Tevatron Run II. Similar conclusions
about the inclusive production at the LHC were obtained in 
Ref.~\cite{Catani:2001ic}.
The reliability of these NNLO approximations follows from the fact that
a sizeable part of the higher-order QCD corrections to the partonic
cross section is due to the emission of soft radiation, whereas the
effect of hard radiation is suppressed by the steeply falling behaviour of the
parton distributions at large $x$. As is customary practice, the perturbative
expansion has been performed in the \ms\ factorization scheme.
Since the \ms-scheme parton densities embody (by definition) too 
much soft-gluon suppression (see e.g. Ref.~\cite{Catani:1997vp}),
the perturbative corrections to the  
partonic cross section have to compensate for it; they thus enhance the
hadronic cross section. At the Tevatron, we have shown that 
the NNLO effect is large and 
increases the cross
section by about $50\%$ with respect to the NLO result.
This suggest
that contributions beyond the NNLO can still be significant.
Since Higgs boson production is less close to threshold at the LHC than at
the Tevatron, the accompanying QCD radiation is less soft  
at the LHC than at the Tevatron. The estimated size of the NNLO effects at the 
LHC \cite{Catani:2001ic} is thus consistently smaller than at the Tevatron.

We have then addressed the impact of a jet veto on the inclusive cross section,
both at the Tevatron and at the LHC. We have presented results at NLO and NNLO.
At NNLO, we have used our estimate for the inclusive cross section 
and we have computed (subtracted) the effect of the jet veto by using the 
numerical program of Ref.~\cite{deFlorian:1999zd}. The jet veto reduces not 
only the absolute value of the cross section, but also the size of the
higher-order QCD corrections.
Since the accompanying
QCD radiation is softer at the Tevatron than at the LHC,
the impact of the veto is less effective at the Tevatron than at the LHC.
In the case of a strong transverse-momentum cut of $p_T^{\rm veto}=15$~GeV
at the Tevatron, the NNLO contributions increase the NLO result by
about $25\%$. At the LHC, the difference between the NNLO and NLO results
is already reduced to less than $10\%$ by considering a weaker cut of 
$p_T^{\rm veto}=30$~GeV. By further decreasing $p_T^{\rm veto}$, perturbative
contributions beyond the NNLO can become sizeable, since they are enhanced
by powers of logarithmic terms of the type $\ln p_T^{\rm veto}$.

In this paper we have limited our analysis to the SM Higgs boson.
The NLO QCD corrections to Higgs boson production within 
the Minimal Supersymmetric extension of the Standard Model (MSSM)
are known \cite{Spira:1995rr,Dawson:1996xz,Spira:1998dg}; 
for small values $(\tan \beta \ltap 5)$ of the MSSM parameter $\tan \beta$,
they are comparable to those for the SM Higgs. This suggests that this
similarity of the size of the perturbative QCD corrections remains true also 
at NNLO.

In actual experiments at hadron colliders, the observed cross section for
Higgs signal events
can significantly depend on the details of the experimental analysis.
This is particularly true when a jet veto procedure is applied, because of
effects due to efficiency for jet reconstruction, jet energy
calibration, presence of pile-up events and so forth. Quantitative
estimates
of these effects require event simulations
\cite{Carena:2000yx,atlascms} based on parton shower Monte Carlo
generators
\cite{Marchesini:1992ch,Sjostrand:2000wi}. The parton shower produces
multi-parton configurations that approximate the exact QCD matrix
elements,
but such approximation does not strictly correspond to the perturbative
expansion of the cross section at LO, NLO, NNLO and so forth.
Therefore, the NLO or NNLO $K$-factors computed in this paper
cannot naively be used to rescale the results of present
Monte Carlo simulations at the the Tevatron and the LHC.
As is customary practice in QCD analyses, more refined studies, which
combine
the perturbative QCD predictions with the Monte Carlo simulations at the
detector level, are necessary to firmly quantify the
expected number of veto selected Higgs events at Tevatron and the LHC.

\vspace{1cm}
\noindent {\bf Acknowledgements.} 
Part of this work was performed during the Workshop on `Physics at TeV
Colliders' (Les Houches, France, May 2001).
We would like to thank Elzbieta Richter-Was for many useful discussions
and comments. 
We also thank Thomas Gehrmann and Stefano Moretti for comments on the draft.


\begin{thebibliography}{90}


\bibitem{Gunion:1989we}
For a review on Higgs physics in and beyond the Standard Model, see
J.~F.~Gunion, H.~E.~Haber, G.~L.~Kane and S.~Dawson,
{\it The Higgs Hunter's Guide} (Addison-Wesley, Reading, Mass., 1990).

\bibitem{lep}
The LEP Collaborations and the LEP Working Group for Higgs boson searches,
hep-ex/0107029. 

\bibitem{lepc}
P.~Igo-Kemenes, presentation given at the open session of the LEP Experiments 
Committee Meeting, 3~November 2000 (see:
http://lepHiggs.web.cern.ch/LEPHIGGS/talks/index.html).

\bibitem{leppapers}
R.~Barate {\it et al.}, ALEPH Coll.,
Phys.\ Lett.\ B {\bf 495} (2000) 1;
M.~Acciarri {\it et al.}, L3 Coll.,
Phys.\ Lett.\ B {\bf 495} (2000) 18;
P.~Abreu {\it et al.}, DELPHI Coll.,
Phys.\ Lett.\ B {\bf 499} (2001) 23;
G.~Abbiendi {\it et al.}, OPAL Coll.,
Phys.\ Lett.\ B {\bf 499} (2001) 38;
P.~Achard {\it et al.}, L3 Coll.,
Phys.\ Lett.\ B {\bf 517} (2001) 319.

\bibitem{ew}
The LEP Collaborations, the LEP Electroweak Working Group and the SLD Heavy 
Flavour and Electroweak Working Group, CERN report LEPEWWG/2001-01.


\bibitem{Carena:2000yx}
M.~Carena {\it et al.},
Report of the Tevatron Higgs Working Group,
hep-ph/0010338.

\bibitem{atlascms}
CMS Coll., Technical Proposal, report CERN/LHCC/94-38 (1994);
ATLAS Coll., ATLAS Detector and Physics Performance: Technical Design
Report, Volume 2, report CERN/LHCC/99-15 (1999).

\bibitem{Georgi:1978gs}
H.~M.~Georgi, S.~L.~Glashow, M.~E.~Machacek and D.~V.~Nanopoulos,
Phys.\ Rev.\ Lett.\  {\bf 40} (1978) 692.

\bibitem{Spira:1998dg}
M.~Spira,
Fortsch.\ Phys.\ {\bf 46} (1998) 203.

\bibitem{angcor}
C.~A.~Nelson,
Phys.\ Rev.\ D {\bf 37} (1988) 1220;
M.~Dittmar and H.~Dreiner,
Phys.\ Rev.\ D {\bf 55} (1997) 167.

\bibitem{Han:1999ma}
T.~Han and R.~Zhang,
Phys.\ Rev.\ Lett.\  {\bf 82} (1999) 25;
T.~Han, A.~S.~Turcot and R.~Zhang,
Phys.\ Rev.\ D {\bf 59} (1999) 093001.

\bibitem{Dawson:1991zj}
S.~Dawson,
Nucl.\ Phys.\ B {\bf 359} (1991) 283.

\bibitem{Djouadi:1991tk}
A.~Djouadi, M.~Spira and P.~M.~Zerwas,
Phys.\ Lett.\ {\bf B264} (1991) 440.

\bibitem{Spira:1995rr}
M.~Spira, A.~Djouadi, D.~Graudenz and P.~M.~Zerwas,
Nucl.\ Phys.\ {\bf B453} (1995) 17.

\bibitem{Kramer:1998iq}
M.~Kramer, E.~Laenen and M.~Spira,
Nucl.\ Phys.\ B {\bf 511} (1998) 523.

\bibitem{Catani:2001ic}
S.~Catani, D.~de Florian and M.~Grazzini,
JHEP {\bf 0105} (2001) 025.


\bibitem{Harlander:2001is}
R.~V.~Harlander and W.~B.~Kilgore,
Phys.\ Rev.\ D {\bf 64} (2001) 013015.


\bibitem{Harlander:2000mg}
R.~V.~Harlander,
Phys.\ Lett.\ {\bf B492} (2000) 74.


\bibitem{Campbell:1998hg}
J.~M.~Campbell and E.~W.~Glover,
Nucl.\ Phys.\ B {\bf 527} (1998) 264.

\bibitem{Catani:2000ss}
S.~Catani and M.~Grazzini,
Nucl.\ Phys.\ {\bf B570} (2000) 287.

\bibitem{Bern:1998sc}
Z.~Bern, V.~Del Duca and C.~R.~Schmidt,
Phys.\ Lett.\ {\bf B445} (1998) 168;
Z.~Bern, V.~Del Duca, W.~B.~Kilgore and C.~R.~Schmidt,
Phys.\ Rev.\ {\bf D 60} (1999) 116001.

\bibitem{Catani:2000pi}
S.~Catani and M.~Grazzini,
Nucl.\ Phys.\ {\bf B591} (2000) 435.

\bibitem{Matsuura:1989sm}
T.~Matsuura, S.~C.~van der Marck and W.~L.~van Neerven,
Nucl.\ Phys.\ {\bf B319} (1989) 570.

\bibitem{mrst2000}
A. D. Martin, R. G. Roberts, W. J. Stirling and R. S. Thorne,
Eur. Phys. J. {\bf C18} (2000) 117.

\bibitem{vnvogt}
W.~L.~van Neerven and A.~Vogt,
Phys.\ Lett.\ {\bf B490} (2000) 111,
Nucl.\ Phys.\ {\bf B588} (2000) 345.

\bibitem{moriond}
S.~Catani, D.~de Florian and M.~Grazzini,
preprint CERN-TH/2001-147 [hep-ph/0106049],
presented at the 36th Rencontres de Moriond on QCD and Hadronic Interactions, Les Arcs, France, 17-24 March 2001 and
at the 9th International Workshop on Deep Inelastic Scattering (DIS 2001),
Bologna, Italy, 27 April - 1 May 2001.

\bibitem{deFlorian:1999zd}
D.~de Florian, M.~Grazzini and Z.~Kunszt,
Phys.\ Rev.\ Lett.\  {\bf 82} (1999) 5209.


\bibitem{efflag}
J.~Ellis, M.K.~Gaillard and D.V.~Nanopoulos, 
Nucl.\ Phys.\ {\bf B106} (1976) 292;
A. Vainshtein, M. Voloshin, V. Zakharov and M. Shifman,
Sov. J. Nucl. Phys. {\bf 30} (1979) 711
[Yad.\ Fiz.\  {\bf 30} (1979) 1368].

\bibitem{Chetyrkin:1997iv}
K.~G.~Chetyrkin, B.~A.~Kniehl and M.~Steinhauser,
Phys.\ Rev.\ Lett.\  {\bf 79} (1997) 353,
Nucl.\ Phys.\ B {\bf 510} (1998) 61.

\bibitem{Contogouris:1990zq}
A.~P.~Contogouris, N.~Mebarki and S.~Papadopoulos,
Int.\ J.\ Mod.\ Phys.\ A {\bf 5} (1990) 1951.


\bibitem{Gluck:1998xa}
M.~Gluck, E.~Reya and A.~Vogt,
Eur.\ Phys.\ J.\ {\bf C5} (1998) 461.

\bibitem{cteq5}
H.~L.~Lai {\it et al.}  [CTEQ Coll.],
Eur.\ Phys.\ J.\ C {\bf 12}, 375 (2000).

\bibitem{Huston:1998jj}
J.~Huston, S.~Kuhlmann, H.~L.~Lai, F.~Olness, J.~F.~Owens, D.~E.~Soper and W.~K.~Tung,
Phys.\ Rev.\ D {\bf 58} (1998) 114034.

\bibitem{Harlander:2001eb}
R.~V.~Harlander and W.~B.~Kilgore,
preprint BNL-HET-01-36 [hep-ph/0110200].

\bibitem{Schmidt:1997wr}
C.~R.~Schmidt,
Phys.\ Lett.\ B {\bf 413} (1997) 391.

\bibitem{Dawson:1992au}
S.~Dawson and R.~P.~Kauffman,
Phys.\ Rev.\ Lett.\  {\bf 68} (1992) 2273.
R.~P.~Kauffman, S.~V.~Desai and D.~Risal,
Phys.\ Rev.\ D {\bf 55} (1997) 4005
[Erratum {\it ibid.}\ D {\bf 58} (1997) 119901].

\bibitem{Ellis:1988xu}
R.~K.~Ellis, I.~Hinchliffe, M.~Soldate and J.~J.~van der Bij,
Nucl.\ Phys.\ B {\bf 297} (1988) 221;
U.~Baur and E.~W.~Glover,
Nucl.\ Phys.\ B {\bf 339} (1990) 38.

\bibitem{DelDuca:2001eu}
V.~Del Duca, W.~Kilgore, C.~Oleari, C.~Schmidt and D.~Zeppenfeld,
preprint  MADPH-01-1226 [hep-ph/0105129],
preprint MADPH-01-1235 [hep-ph/0108030].

\bibitem{Frixione:1996ms}
S.~Frixione, Z.~Kunszt and A.~Signer,
Nucl.\ Phys.\ B {\bf 467} (1996) 399.

\bibitem{Catani:1997vp}
S.~Catani,
hep-ph/9709503,
in Proc. of the 32nd Rencontres de Moriond: QCD and High-Energy
Hadronic Interactions, ed. J. Tran Than Van (Editions Fronti\`eres, 
Gif-sur-Yvette, 1997) p.~331. 

\bibitem{Dawson:1996xz}
S.~Dawson, A.~Djouadi and M.~Spira,
Phys.\ Rev.\ Lett.\  {\bf 77} (1996) 16.


\bibitem{Marchesini:1992ch}
G.~Marchesini, B.~R.~Webber, G.~Abbiendi, I.~G.~Knowles, M.~H.~Seymour
and L.~Stanco,
Comput.\ Phys.\ Commun.\  {\bf 67} (1992) 465;
G.~Corcella, I.G.~Knowles, G.~Marchesini, S.~Moretti, K.~Odagiri,
P.~Richardson, M.H.~Seymour and B.R.~Webber,
JHEP {\bf 0101} (2001) 010.


\bibitem{Sjostrand:2000wi}
T.~Sjostrand, P.~Eden, C.~Friberg, L.~Lonnblad, G.~Miu, S.~Mrenna and
E.~Norrbin,
Comput.\ Phys.\ Commun.\  {\bf 135} (2001) 238.






\end{thebibliography}
\end{document}